\documentstyle[12pt]{article}
\newcommand{\beqnst}{\begin{eqnarray*}}
\newcommand{\eeqnst}{\end{eqnarray*}}
\newcommand{\beqn}{\begin{eqnarray}}
\newcommand{\eeqn}{\end{eqnarray}}

\begin{document}

\title{\bf Reduction of  Anyons to One Dimension and 
Calogero-Sutherland-type Models}

\author{ Radhika Vathsan \thanks{ email: radhika@imsc.ernet.in.}\\
        {\it The Institute of Mathematical Sciences},\\
{\small C.I.T. Campus, Taramani, Chennai 600 113, INDIA.}}






\date{September 3 !997}
\maketitle 

\begin{abstract}
 The two-dimensional anyon system, when reduced to one dimension,
yields models related to the Calogero-Sutherland model. One such
reduction leads to a new
model with a class of  exact solutions. This model
is one of a family of models obtained upon dimensional reduction
of spherically symmetric models in arbitrary dimensions.

\end{abstract}

\clearpage
\section{Introduction}

Anyons \cite{anyons}, as is well known now, are particles in 
two space dimensions 
with fractional exchange  statistics. The anyon system is an
 example of quantum mechanics on
multiply connected spaces and  has wave functions that
are multi-valued. A gauge transformation can be applied to trade this
multi-valuedness for a specific potential, the so-called
``statistical interaction". Thus a system of anyons can be treated
as interacting particles (e.g. fermions ) with single-valued wave
functions.

In one dimension, the Calogero-Sutherland model
model (\cite{CM},\cite{SM}) is an example of a system exhibiting
fractional exclusion statistics \cite{Pasq}, in the sense
proposed by Haldane
\cite{Hal}.
There has been considerable speculation and interest in the connection
 between the anyon system and the Calogero-Sutherland
(referred to henceforth as CSM), especially in regard to 
the connection between
exchange and exclusion statistics 
(see, for instance,\cite{EXS}).
Ideas of fractional exchange statistics in one dimension have
been studied \cite{Poly}, and so have field theoretic models
\cite{Field}.
Dimensional reduction of the anyon system 
is a way of studying this connection \cite{HLM}.
These works are aimed at understanding the nature of statistics in one
dimension and the approach taken is to arrive at a
one-dimensional model with anyonic features, taking hints
from the two-dimensional case of anyons.

Reducing the dimensions of the anyon system 
model could bear interesting relations  to the CSM. 
Also, one could examine the possibility of obtaining
exact solutions for the N-anyon problem (the non-linearly
interpolating solutions) from possible integrable models obtained 
by such  reduction. 

The present work takes  a different approach to the process of 
``reduction", by which is meant reducing the number of dimensions
in the {\it eigenvalue equation } for the Hamiltonian of the system.
This way, one ensures that the wave functions themselves reduce to
those of the derived model. 
The main motivation behind this is that since we are within the realms
of the original model itself, we can now try to deduce exact solutions for
the two-dimensional anyon system, in  the non-linearly
interpolating regime.
As it turns out, this cannot be achieved because of the peculiar
restrictions under which such a dimensional reduction is possible.

Reduction to one dimension can be carried out in two ways:  in
polar coordinates, one can either freeze the radial coordinates
to arrrive at a model defined on a circle, or freeze the angular
coordinates to get a model on a (half-)line.
In this paper we examine both these cases.

We first set the stage in section 2, describing the anyon system 
and fixing notations.

The case of reduction to the circle forms section 3. This has
been considered in Ref.\cite{BL} but the
resulting model is not 
a reduced anyon model. They arrive at the  Sutherland model 
after a naive reduction by fixing one set of
coordinates. 
However, to ensure that one is still
within the realms of
the same physical model, one must impose the restriction of
coordinates on the whole eigenvalue equation, i.e on the wave
functions {\em after}
the derivatives are evaluated. Then one finds that the original
eigenvalue equation ceases to be an eigenvalue
equation. Impositions of restrictions on the radial behaviour of
the wave functions yields inhomogeneous terms in the equation,
that must be dealt with in order that the reduced equation
become
an eigenvalue equation.

The reduction to a line is considered in  section 4.
Here, direct reduction yields a model which has no solutions
derivable from the anyon system because of the restriction on
the angular dependence of the wave functions. However, this
model is of interest in that it can be obtained  by the  reduction
of a spherically symmetric problem in arbitrary dimensions, with
three-body interactions of the CSM type.
In restricting to models with anyonic solutions, we find that we
must first extract the factor
$\Pi_i \rho_{i}^{-1/2}$ which is the characteristic of anyonic
behaviour and then proceed to fix the angular coordinates. In
this case we obtain a model related to the CSM which has a class
of exact solutions derived from the original anyon solutions.

We conclude with
a summary and discussion of results.

\section{ The Anyon System}
For N anyons in a plane in an oscillator potential, the Hamiltonian is,
 in terms of dimensionless quantities,
\beqn
H = \frac{1}{2} \sum_{i=1}^{N} \left[ ({\bf p}_i - {\bf a}_i)^2 
+ {\bf r}_{i}^2 \right]
\eeqn
 where the statistics, contained in the parameter $\alpha$,
 has been incorporated in terms of a
singular gauge potential
\[ {\bf a}_i = \alpha \hat{\bf z} \times 
\sum_{j \neq i} \frac{{\bf r}_{ij}}{r_{ij}^2}. \]
(The system is confined by an  oscillator potential in order 
ensure  a discrete spectrum.)
Putting in ${\bf p}_i = - \iota{\bf \nabla}_i$ and expanding ,
\beqn
H = -\frac{1}{2}\sum_{i=1}^{N} \left( {\nabla}_{i}^2 - r_{i}^2 \right)
    + \iota \alpha \sum_{i<j=1}^N \frac{l_{ij}}{r_{ij}^2}
    + \frac{\alpha^2}{2} \sum_{i \neq j,k=1}^N 
     \frac{{\bf r}_{ij}.{\bf r}_{ik}}{r_{ij}^2 r_{ik}^2}
 \label{anyon1}
\eeqn
where \( l_{ij} = {\bf r}_{ij} \times {\bf \nabla}_{ij}. \)

For our purpose it is more convenient to work in polar coordinates,

\[ x_i + \iota y_i = \rho_{i} e^{\iota \theta_{i}},
\;\;   \partial_{i} = \frac{1}{2} e^{-\iota \theta_{i}} 
                  \left( \frac{\partial}{\partial \rho_{i}}
                  - \frac{\iota}{\rho_{i}}\frac{\partial}{\partial \theta_{i}}
		  \right)  , \]
so that
\beqn
H&=&-\frac{1}{2} \sum_{i=1}^N \left( \frac{\partial^2}{\partial {\rho}_i^2}
     + \frac{1}{\rho_i} \frac{\partial}{\partial \rho_i} 
     + \frac{1}{\rho_i^2} \frac{\partial^2}{\partial {\theta}_i^2} 
     - \rho_i^2 \right) 
 \nonumber \samepage \\
  &&- \frac{\alpha}{2} \sum_{i<j} 
\left[
   \frac{ e^{-\iota \theta_i}
       \left( 
          \frac{\partial}{\partial \rho_i}
        - \frac{\iota}{\rho_i}\frac{\partial}{\partial \theta_i}
       \right)
        - e^{-\iota \theta_j}
       \left( 
          \frac{\partial}{\partial \rho_j} 
        - \frac{\iota}{\rho_j}\frac{\partial}{\partial \theta_j} 
       \right)}
    {\rho_i e^{-\iota \theta_i} - \rho_j e^{-\iota \theta_j} }
\right.
\nonumber \samepage \\
&&\;\;\;\;\;\;\;\;\;\;\;\; 
\left.
 - \frac{ e^{\iota \theta_i} 
        \left( 
          \frac{\partial}{\partial \rho_i} +
\frac{\iota}{\rho_i}\frac{\partial}{\partial \theta_i}
        \right) 
       - e^{\iota \theta_j}
        \left(  
              \frac{\partial}{\partial \rho_j} +
\frac{\iota}{\rho_j}\frac{\partial}{\partial \theta_j}
        \right)}    
       {\rho_i e^{\iota \theta_i} - \rho_j e^{\iota \theta_j}}
\right] \nonumber \samepage \\
     &&+\frac{\alpha^2}{2} \sum_{i \neq j,k}
         \frac{1}{\left( \rho_i e^{-\iota \theta_i} - \rho_j e^{-\iota
\theta_j} \right) \left( \rho_i e^{\iota \theta_i} - \rho_k e^{\iota
\theta_k} \right)}. \label{anyon}
\eeqn
The eigenvalue equation is 
\beqn
H \Psi( \rho_i, \theta_i) = E \Psi( \rho_i, \theta_i).
\label{eigenvalue}
\eeqn
The proposed ``reduction" of dimensions is carried out at the
level of this eigenvalue equation.
This can be achieved naively in two possible ways:
(a) by fixing the radial variables $\rho_i$ so
that the equation depends only on the $\theta_i$'s and the problem reduces
to particles on a circle of fixed radius $\rho$ and
(b) by keeping the angular variables $\theta_i$ constant, in
which case the problem is only $\rho_i$-dependent and looks like
particles on a line. 
We proceed to examine both these methods now.

\section{Reduction of  Free Anyons to a Circle}
Consider the eigenvalue equation for free anyons 
(no oscillator potential)
at \linebreak $\rho_i = \rho \; \forall \;i $. With no approximations whatever,
we get 
\beqn
H\Psi &=&-\frac{1}{2}
 \left. \sum_{i=1}^N \frac{\partial^2 \Psi}{\partial \rho_i^2}
 \right|_{\rho_i = \rho}  
       -\frac{1}{2}
 \left.  \sum_{i=1}^N \frac{1}{\rho_i}\frac{\partial \Psi}{\partial \rho_i} 
\right|_{\rho_i = \rho}
\nonumber \\
       &&-\left.\frac{\alpha}{2} \sum_{i<j}
         \left[ \frac{1}{\rho_i e^{-\iota
\theta_i} - \rho_j e^{-\iota \theta_j} }
\left(e^{-\iota \theta_i}\frac{\partial}{\partial \rho_i} - e^{-\iota
\theta_j}\frac{\partial}{\partial \rho_j}\right) \right.
\right.
\nonumber \\
& & \left.\;\;\;\;\;\;\;\;\;\;\;\;\;\; \left.-\frac{1}{\rho_i e^{\iota
\theta_i} - \rho_j e^{\iota \theta_j} }
\left(e^{\iota \theta_i}\frac{\partial}{\partial \rho_i} - e^{\iota
\theta_j}\frac{\partial}{\partial \rho_j}\right)\right]\Psi 
\right|_{\rho_i = \rho_j = \rho}
\nonumber \\
 &&- \frac{1}{2 \rho^2}\sum_i \frac{\partial^2\Psi}{\partial
\theta_i^2} + \frac{\iota \alpha}{2\rho^2}(N-1) \sum_i
\frac{\partial\Psi}{\partial \theta_i}
\nonumber \\
 &&+ \frac{\alpha^2}{8 \rho^2}\sum_{i \neq j}
\frac{1}{\sin^2(\frac{\theta_{ij}}{2})}\Psi
 + \frac{\alpha^2}{6\rho^2}N(N-1)(N-2)\Psi. \label{circleq}
\eeqn
We wish to fix the radial coordinates so that the dynamics
depends only on the angular variables and the problem is reduced
to one dimension, {\em viz} the circle.
To achieve this, one could naively drop the $\rho$-derivatives
so that only the last two lines in the above equation survive and the
Hamiltonian becomes, after some regrouping,
\beqn
H'= \frac{1}{2\rho^2} 
    \left\{
         \sum_{i=1}^N \left[ \iota \frac{\partial}{\partial \theta_i} +
\alpha(N-1)\right]^2
    + \frac{\alpha^2}{4}\sum_{i \neq j}
\frac{1}{\sin^2(\frac{\theta_{ij}}{2})}
    - \frac{\alpha^2}{3}N(N-1)(2N-1)
    \right\}.
\eeqn
This is the Sutherland model, which is related to the Calogero model
on a circle, but for the shifted momentum operator and the constant
$\alpha^2$-dependent term that will just add to the eigenvalue. 
The shift can be annulled by performing a gauge transformation
 on the wave function:
\[ \Psi \rightarrow \Psi \exp( \iota \alpha(N-1)\sum_i \theta_i). \]
This is the  reduction obtained by Bhaduri and Li \cite{BL}. 
However, this model is {\em not} a reduced anyon model, because
it rests on the over-restrictive condition that the wave functions be
 completely $\rho$-independent. 
 (Nevertheless it could exist on its own right
as some one-dimensional model, albeit with no {\em anyonic}
connection!)

If we wish to retain the connection with the anyon system
in such a dimensional reduction,
the correct way to proceed is to evaluate the
derivatives first and then equate all the radial variables
to a fixed value $\rho$. Let us see how  
Eq (\ref{circleq}) can be simplified.
Now the  $\rho_i$-derivatives in the $\alpha$-dependent term in
 will
cancel if we consider the rather special case of wave functions
that factor into radial and angular parts and
with the radial part symmetric in the $\rho_i$'s.
In such a case, the Hamiltonian will look like
\beqn
H_{red}&=&  -\frac{N}{2}
         \left[  \frac{\partial^2}{\partial \rho^2}
         + \frac{1}{\rho} \frac{\partial}{\partial \rho}
	 \right]
	 + H'.
\eeqn
Thus the eigenvalue equation for this Hamiltonian is not
that of the Sutherland model, but has inhomogeneous
terms arising from the $\rho$-derivatives of the wave
function.  One could get rid of these terms in several ways:
(a) One could demand that $\rho$-derivatives
cancel each other. This could happen by choosing $\rho_i$-dependence
of the wave functions to be of the form $\ln\rho_i$, appropriately
symmetrised. However these may not be solutions to the original anyon
problem.
(b) One could evaluate  the inhomogeneous terms
for a fixed $\rho$ and choose only that value (if any) where
these terms vanish.  
(c) Another possibility is to choose boundary conditions such that
the contributions to the $\rho$-derivatives at the boundary
cancel against the extra terms.

Suffice it to say that without additional restrictions, one does
not obtain the Sutherland model or a related model on trying to 
restrict the anyon
system to a circle.

\section{Reduction to a Line}

Another way of reducing the dimensions in the Hamiltonian 
(\ref{anyon}) is to fix the angular variables and hence constrain
the problem to a line. 
This also means that we wish to project out $\theta$-independent
solutions of the anyon equation (\ref{anyon1}). 
Now the singular nature of the potential at $r_i = r_j$ demands
that well-behaved wave functions be of the form
$\Pi_{i<j}|r_{ij}|^\alpha \psi(r_i)$.
Here we must look at the eigenvalue equation
when all the $\theta_i$'s are equated
after evaluating the angular derivatives on the wave function.
Since the total angular momentum
$ J = \iota \sum_{i=1}^N \frac{\partial}{\partial \theta_i} $
is conserved, we can look at solutions for which 
the angular dependence of the wave function can be
separated as
\beqn
\Psi( \rho_i, \theta_i) = \sum_{\{n_i\}} \exp(\iota \Sigma_i n_i \theta_i)
\psi_{n_i}(\rho_i), \label{theta}
\eeqn
where $ \sum_i n_i = j $ is the total angular momentum eigenvalue.
The  eigenvalue equation becomes
\beqn
\sum_{\{n_i\}} H_{n_i} \psi_{n_i}(\rho_i) =  E \sum_{\{n_i\}}\psi_{n_i}(\rho_i), 
\eeqn
where
\beqn
H_{n_i}
&=& 
  -\frac{1}{2} \sum_{i=1}^N
        \left( \frac{\partial^2}{\partial\rho_i^2}
             + \frac{1}{\rho_i} \frac{\partial}{\partial\rho_i}
             - \frac{n_i^2}{\rho_i^2}
        \right) 
   + \frac{1}{2}\sum_{i=1}^N \rho_{i}^2 
  \nonumber \samepage \\
&& - \frac{\alpha}{2} \sum_{i<j} 
   \left[ \frac{ e^{-\iota \theta_i} 
                \left( \frac{\partial}{\partial\rho_i} + \frac{n_i}{\rho_i}
                \right)
                - e^{-\iota \theta_j} 
                \left( \frac{\partial}{\partial\rho_j} + \frac{n_j}{\rho_j}
                \right)
		}
          {\rho_i e^{-\iota \theta_i} - \rho_j e^{-\iota \theta_j} }
   \right.
\nonumber \samepage \\
&&
\left. \;\;\;\;\;\;\;\;\;\;\;\;
         - \frac{ e^{\iota \theta_i} 
                 \left(  \frac{\partial}{\partial\rho_i} - \frac{n_i}{\rho_i}
        	 \right)   
       - e^{\iota \theta_j}
       		 \left( \frac{\partial}{\partial\rho_j} - \frac{n_j}{\rho_j}  
	         \right)
		}
          {\rho_i e^{\iota \theta_i} - \rho_j e^{\iota \theta_j}}
    \right] \nonumber \samepage \\
  && + \frac{\alpha^2}{2} \sum_{i \neq j,k}
    \frac{1}{ ( \rho_i e^{-\iota \theta_i} - \rho_j e^{-\iota \theta_j} )
              ( \rho_i e^{\iota \theta_i} - \rho_k e^{\iota \theta_k} )
   	    }.
\eeqn
In contrast to the previous case, we have gotten rid of the
$\theta$-derivatives without the complication of inhomogeneous
terms.
But now the eigenvalue equation involves
a sum over all possible sets $\{n_i: \Sigma_i n_i = j\}$ 
and in complete generality, involves considering
infinite sets of $n_i$'s. The problem becomes hard to tackle
because of this.
If, for simplicity, we choose 
 just one set, the the sum collapses to one term.
(We must keep in mind the restrictions imposed by this choice.)
On equating all the angular variables, we get the Hamiltonian for
anyons restricted to  a line as 
\beqn
H_{n_i} &=& - \frac{1}{2} \sum_{i=1}^N \left( \frac{\partial^2}{\partial\rho_{i}^2} +
\frac{1}{\rho_i}\frac{\partial}{\partial\rho_{i}} - \frac{n_i^2}{\rho_i^2} \right)
  +\frac{1}{2} \sum_{i=1}^N \rho_i^2
\nonumber \samepage \\
&&  - 2\alpha \sum_{i<j} \frac{n_i \rho_j - n_j \rho_i}
                         {\rho_j \rho_i (\rho_i - \rho_j)}
  + \frac{\alpha^2}{2} \sum_{i \neq j} \frac{1}{(\rho_i - \rho_j)^2}.
\eeqn

If we now choose $ n_i = 0 \:\forall \: i$, the Hamiltonian
simplifies to that of the standard CSM but for the
$\frac{1}{\rho_i}\frac{\partial}{\partial\rho_{i}}$ term.
This can easily be brought to the standard CSM form 
by taking 
$ \psi_0 (\rho_i) = \prod_i \rho_{i}^{-1/2} \chi(\rho_i)$.
 This scaling is also consistent with that required of the 
inner product when reducing the dimensions to one, i.e ensures
that 
\( \int\rho d\rho |\psi_0|^2 \rightarrow \int d\rho |\chi|^2 \). 
 In the process, however, an extra potential is generated.
$\chi(\rho_i)$ is then an eigenfunction of the 
Hamiltonian
\beqn
 H_{red} = - \frac{1}{2} \sum_i \frac {\partial^2}{\partial\rho_{i}^2}
 + \frac{\alpha^2}{2} \sum_{i \neq j}
\frac{1}{(\rho_i - \rho_j)^2} 
 + \frac{1}{2} \sum_i \left( \rho_i^2 - \frac{1}{4\rho_i^2} 
\right)\label{model}
\eeqn
which is the CSM Hamiltonian \cite{CM} (in an oscillator
confinement) but with two noteworthy differences:
(a) the additional attractive $\frac{1}{8\rho_i^2}$ potential.
(b) the range of the variable $\rho_i$ being ($ 0 ,
\infty$), this model is defined on the half-line.

When seeking solutions to this model from those of the original
anyon system,
we notice that the wave functions,
that contain the factor 
$\Pi_i \rho_{i}^{-1/2}$,
can no longer 
be factorised as in (\ref{theta}) for a finite sum over the
$n_i$'s.
Therefore this model no longer has a direct bearing on the anyon system.
To stay within the purview of anyonic solutions, one must first
take out this factor and then reduce.
This  will be explored
in the sub-section that follows.

However, the model (\ref{model}) is interesting in its own right.
Note that a CSM model
modified by a term of the form
$\frac{a}{\rho_i}\frac{\partial}{\partial\rho_{i}}$
can be brought to the standard CSM form by redefining the wave
function. Now such a term appears in the radial part of the
Laplacian in $d$ dimensions:
\[ {\bf \nabla}^2 = \frac{\partial^2}{\partial \rho^2} +
\frac{d-1}{\rho} \frac{\partial}{\partial \rho} +
{\bf \nabla}_{\perp}^2 \]
where ${\bf \nabla}_{\perp}$ is the  part of the
Laplacian orthogonal to the radial part.
On pulling out a factor $ \rho^{-d/2}$ from the wave
function, the radial part reduces to 
\[{\bf \nabla}^2_{\rho} =  \frac{\partial^2}{\partial \rho^2}
 - \frac{(d-1)(d-3)}{4 \rho^2}, \]
generating an extra potential. This scaling is also required
of the inner product when reducing the dimensions. Note that the
extra potential term is attractive for $d = 2$, {\em absent} for $d = 3$
and repulsive for higher dimensions. Thus if our model is
spherically symmetric, then the spherically invariant
eigenfunctions satisfy the eigenvalue equation
\[ -\frac{1}{2}{\bf \nabla}^2_{\rho}\psi(\rho) = E \psi(\rho).\]

Now a potential of the kind $ V({\bf r}_i) = \sum_{i \neq j,k=1}^N
     \frac{{\bf r}_{ij}.{\bf r}_{ik}}{r_{ij}^2 r_{ik}^2}$, when
restricted to a line (fix all angular variables) yields the
standard CSM inverse-square two-body interaction. Therefore we
see that there is a family of higher dimensional models that can
be reduced to a CSM model with a 
$\frac{1}{\rho_i^2}$ potential defined for $ \rho > 0$, that is
therefore important to analyse.

\subsection{Anyon Solutions and a Related Model}

We will now explore a way of effecting a reduction while
remaining within the realm of 
solutions of the anyon equation (\ref{anyon1}).

It is well known (see \cite{DKM}) that exact solutions for
  the anyon eigenvalue equation (\ref{anyon})  are of the form
\beqn
 \Psi(\rho_i, \theta_i)  =  \prod_{i<j}|{\bf r}_i -
{\bf r}_j|^\alpha
               \exp{\left(- \frac{1}{2}\sum_i \rho_i^2\right)}
                            \psi(r_i, \theta_i).
\eeqn
But here, the $\theta$-dependence of the wave function cannot be 
factorised as in eq. (\ref{theta}) if as in the previous section
on chooses only one set of $n_i$'s and not the infinite sum.

Now the singular nature of the  anyonic potential 
at ${\bf r}_i = {\bf r}_j$ demands
that well-behaved wave functions be of the form
$\Pi_{i<j}|{\bf r}_{ij}|^\alpha \psi(r_i).$
Therefore, one must carry out the reduction after extracting this
factor from the wave function in eq. (\ref{eigenvalue}).

$\psi(r_i, \theta_i)$ then satisfies the eigenvalue equation of the
Hamiltonian
\beqn
H&=&-\frac{1}{2} \sum_{i=1}^N \left( \frac{\partial^2}{\partial
{\rho}_i^2}
     + \frac{1}{\rho_i} \frac{\partial}{\partial \rho_i}
     + \frac{1}{\rho_i^2} \frac{\partial^2}{\partial {\theta}_i^2}
     - \rho_i^2 \right) \nonumber \\
&&  - \alpha \sum_{i<j}
    \frac{ e^{-\iota \theta_i}
\left( 
          \frac{\partial}{\partial \rho_i}
        - \frac{\iota}{\rho_i}\frac{\partial}{\partial \theta_i}
       \right)
     - e^{-\iota \theta_j}
\left(
          \frac{\partial}{\partial \rho_j}
        - \frac{\iota}{\rho_j}\frac{\partial}{\partial \theta_j}
       \right)
}
    {\rho_i e^{\iota \theta_i} - \rho_j e^{\iota \theta_j}}
\eeqn
 Here if we look for $\theta$-dependence of the form (\ref{theta})
and choose $n_i = 0 \: \forall\:  i$, the $\theta$ dependence drops out and
$\chi(\rho_i)$ satisfies the eigenvalue equation of the Hamiltonian
\beqn
H&=&-\frac{1}{2} \sum_{i=1}^N \left( \frac{\partial^2}{\partial
{\rho}_i^2}   
     + \frac{1}{\rho_i} \frac{\partial}{\partial \rho_i} - \rho_i^2
\right)
 - \alpha \sum_{i<j} \frac{\partial_{ij}}{\rho_{ij}}.
\eeqn
(Here the subscript $ij$ means $ X_{ij} = X_i - X_j$.)
We can now put back the factor $|\rho_{ij}|^{-\alpha}$ (which has no
$\theta$-dependence) into this, and also take out the factor
$\Pi_i \rho_{i}^{-1/2}$
 so that the reduced  Hamiltonian is 
\beqn
H=-\frac{1}{2} \sum_{i=1}^N \frac{\partial^2}{\partial
{\rho}_i^2} +
 \frac{1}{2} \sum_i \left( \rho_i^2 - \frac{1}{4\rho_i^2} \right)
+ \frac{\alpha^2}{2} \sum_{i \neq j}
\frac{1}{(\rho_i - \rho_j)^2}
 +\frac{\alpha}{2} \sum_{i<j}\frac{1}{ \rho_i \rho_j}.
\eeqn
 We now have a model similar to the earlier one (\ref{model}) except for
the new $\alpha$-dependent term. This is a general inverse quadratic
potential in the $\rho$ coordinates, which has solutions obtained from
the anyon model we started out with.
The solutions we need are those corresponding to the condition
$ n_i =0 \: \forall \:i$, which are the $j=0$ solutions.
The wave functions $ \chi_0$ are now functions of the
variable $t = \sum_i \rho_i^2$
and satisfy the confluent hyper-geometric equation (see
\cite{DKM})
with the polynomial solution
\beqn
 \chi_0(t) = M\left(-m, N+\alpha\frac{N(N-1)}{2}, t\right)
\eeqn
 
and normalizability requires that energy is quantized as
\beqn
 E = N + 2m + \alpha\frac{N(N-1)}{2}.
\eeqn

\section{Summary and Conclusion}
 In this work the dimensional reduction of the
N-anyon system to a circle and to a line has been studied from a
mathematical point of view. It must be noted that our philosophy is
slightly different from previous works along these lines, for
instance ref \cite{HLM}, where the physical problem of particles with
exchange statistics in one dimension is dealt with.
Here we take the standard two-dimensional anyon eigenvalue equation
and project it onto one dimension in two ways. The resultant model
not only carries an anyonic flavour but also has solutions projected
out of the two-dimensional anyon solutions.

 Reduction to a circle, if carried out in a naive way by simply dropping the
radial derivatives, leads to the Sutherland model, as obtained
in \cite{BL} (This model seems to arise in connection with edge
states in the composite fermion model for the quantum Hall
effect \cite{YZ}). This process
however does not retain connection with the original anyon
model, since it means imposing restrictions on the
wave functions that are too severe.
 One tries to retain the connections to this exactly
solvable model for some particular class of solutions
to the anyon problem if the radial derivatives are evaluated first and
then chosen so as to cancel. A related model with extra potential
terms can be obtained if the radial terms evaluated at a fixed value
of the radial variables are retained.

Dimensional reduction of the the N-anyon system to a (half-)line
, obtained by requiring the $\theta$-dependence of the wave
functions to factor out and summing over one set out of the
infinite possible sets $n_i$,
 yields a  model  related to the Calogero-Sutherland model.
This model is important from the relation it bears to reduction
of higher dimensional rotationally invariant
models with CSM-like interactions.
However, this model does not have solutions that can be adapted 
from the known exact solutions for anyons. This is because the
factor 
$\Pi_{i<j}|{\bf r}_{ij}|^\alpha $ in anyonic solutions
cannot be factored as in (\ref{theta}) for a single set of
$n_i$s.
So one carries out the reduction {\it
after} taking out this factor. The model now obtained does have
solutions reduced from the anyon solutions. 
These are not the only solutions to the
model 
and means of obtaining other solutions to this problem are under
investigation.

One therefore concludes that the anyon system cannot be directly
reduced (in our sense of the term ``reduction") to one dimension to
give the CSM .

One also concludes that the reduction procedure does not give a way
to deduce solutions to the original anyon model in the regime of
eigenvalues having non-linear dependence on ${\alpha}$.
This is because these solutions require considering infinite sums
over the $n_i$'s (see ref. \cite{DKM}) and here one loses the
connection to the CSM-like model.

\section*{Acknowledgments}
I thank M. V. N. Murthy for suggesting this line of research
and for extremely fruitful discussions. 
The results obtained
here are a product of extensive discussions with G. Date,
 for which I am indebted to him. I am also thankful to both of
them for going through the manuscript carefully and offering
useful suggestions.


\end{document}